\documentclass[a4paper,12pt]{article}
\usepackage{booktabs}%,caption}
\usepackage{booktabs, makecell, longtable}
\usepackage{physics}
\usepackage{rotating}
\usepackage{makecell}
\usepackage[flushleft]{threeparttable}
\usepackage{amssymb}
\usepackage{graphicx}
\usepackage{amsmath}
\usepackage{epstopdf}
\usepackage[hang]{subfigure}
\usepackage{bm}
\usepackage{authblk}
\usepackage[utf8x]{inputenc} 
\usepackage{multicol, multirow}
\usepackage{url}
\usepackage[breaklinks,colorlinks,citecolor=blue,linkcolor=blue]{hyperref} 
\urldef{\mailsa}\path|{silee}@sejong.ac.kr| 

\newcommand{\RNum}[1]{\uppercase\expandafter{\romannumeral #1\relax}}
\newcolumntype{L}[1]{>{\raggedright\arraybackslash}p{#1}}
\newcolumntype{C}[1]{>{\centering\arraybackslash}p{#1}}
\newcolumntype{R}[1]{>{\raggedleft\arraybackslash}p{#1}}
\begin{document}

\title{Hyperparameter Optimization for Forecasting Stock Returns}

\author{Sang Il Lee%
  \thanks{Electronic address: \texttt{sangillee.fin@gmail.com}}}
\affil{DeepAllocation Technologies}

\maketitle

\begin{abstract}
In recent years, hyperparameter optimization (HPO) has become an increasingly important issue in the field of machine learning for the development of more accurate forecasting models. 
In this study, we explore the potential of HPO in modeling stock returns using a deep neural network (DNN). 
The potential of this approach was evaluated using technical indicators and fundamentals examined based on the effect the regularization of dropouts and batch normalization for all input data. 
We found that the model using technical indicators and dropout regularization
significantly outperforms three other models, showing a positive predictability of 0.53$\%$ in-sample and 1.11$\%$ out-of-sample, thereby indicating the possibility of beating the historical average.
We also demonstrate the stability of the model in terms of the changes in its feature importance over time. 

\end{abstract}

\section{Introduction}
\label{intro}
Deep learning has become a promising way to model the complexity of stock movements. It enables us to capture non-linear movements, to associate large data, and to reduce noise without an assumption of a pre-specified underlying structure. At the same time, it leaves us with a difficulty in selecting numerous hyperparameters, which critically affects the performance of the resulting models.
Most studies dealing with a financial time series typically choose pre-specified hyperparameters and check the robustness of the model based on small changes in the parameters. This approach requires experts to put a lot of effort into tuning numerous parameters simultaneously, which often results in a suboptimal model. 

 Hyperparameter optimization (HPO) can be used to mitigate this problem by automatically searching
for the most optimal hyperparameters in machine learning learners, and has been widely used to identify good configurations more quickly, such as through the use of a sequential model-based algorithm configuration (SMAC), tree-structure Parzen estimator (TPE), and Sprearmint \cite{feurer2014using}.
HPO has also been demonstrated to be an extremely powerful approach for automatic image and speech recognition, and offers advantages for dealing with machine learning in a systematic manner. 
First, it reduces the human effort necessary in tuning the hyperparameters and opens up the possibility of improving the performance of machine learning \cite{melis2018state}\cite{snoek2012practical}. Second, it improves the reproducibility and fairness of scientific studies because an automated HPO is more reproducible than a hand-tuned approach using trial-and-error
searches to produce a desired behavior, thereby allowing us to compare different methods more fairly through the same level of tuning \cite{bergstra2013making}\cite{sculley2018winner}.

\begin{figure}[t]
\centering
     \scalebox{0.5}
  {
	\includegraphics{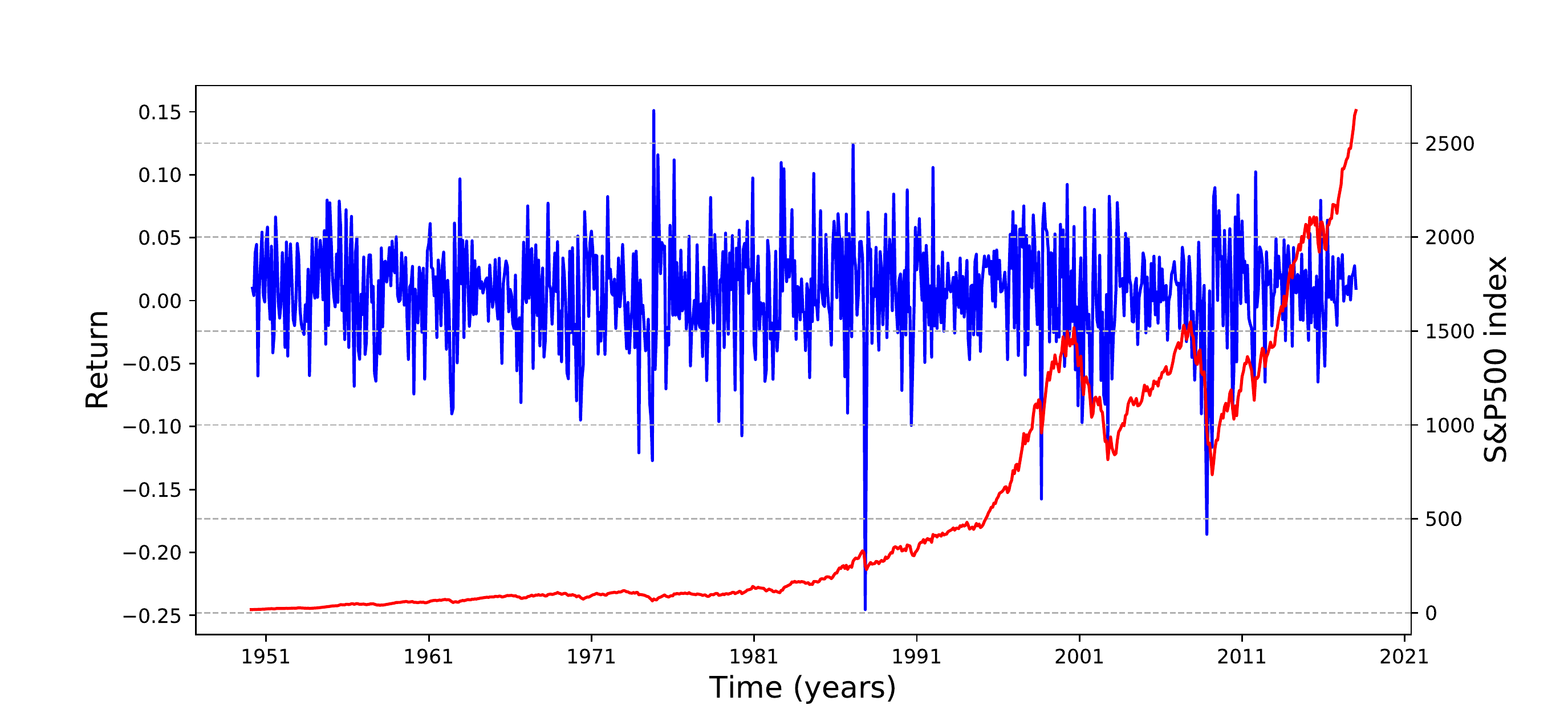}

   }
\caption{S$\&$P 500 index and its returns from Jan. 1, 1950 to Dec. 31, 2017.}
\label{fig:log_return}
\end{figure}

Despite such advantages, financial studies have generally not considered this method. HPO requires a large data scale to avoid an overfitting occurring in both the training and validation data. Stock-related data are obtained only over a relatively short time span, typically from the year 1950 to the present. As shown in Fig. \ref{fig:log_return}, a random evolution of a stock return, such as time-varying volatility and occasional jumps related to crashes or sudden upsurges, causes a time dependency of the model parameter set to specific periods. Furthermore, cross-validation and shuffling, which are crucial techniques for preventing an overfitting, cannot be used because stock-related data are time-ordered, and a modeling process requires preserving the time ordering. 
For these reasons, the use of HPO has
rarely been assessed and there is a poor understanding of its efficiency in financial data modeling. As a result, practitioners need to pay more attention to hyperparameter tuning and the resulting models largely depending on their experience. 

In this study, we evaluate the viability of HPO in terms of the stock return predictability problem. We examined the HPO performance across different conditions, the input features of the fundamentals and technical indicators, and the regularization of a dropout and batch normalization.
Our key findings are as follows:
\begin{itemize}
\item[•] We show that, whereas the prediction models with an input of fundamentals are likely to overfit the in-sample data, 
models with the input feature of the technical indicators achieves a strong predictability throughout the in- and out-of-sample periods. A dropout is more effective for a positive predictability in an out-of-sample than a batch normalization.
\item[•] We show that the model with good predictability in both an in- and out-of-sample is less sensitive to the time evolution, which reveals that it is a general model for adapting to the changes in the economic and business conditions.
\end{itemize}
We believe this study provides insight into the application of machine learning for investment purposes or risk management.
\\
\\
\noindent {\bf Related work} 
 In financial economics, there is a long-standing debate whether (excess) stock market returns are predictable. 
The conventional framework for analyzing equity premium predictability is a `linear predictive regression' model taking the following form:
\begin{equation}
r_{t+1}=\alpha+\bm \beta^{'} \bm x_{t}+\varepsilon_{t+1},
\end{equation}
where $r_{t+1}$ is the return on the stock market index in excess of the risk-free interest rate, $\alpha$ is an intercept term, $\bm \beta$ is a $p\times 1$ dimensional vector of the slope parameters, $\bm x_{t}$ is a $p\times 1$ dimensional vector of the predictor variables observed at time $t$, and $\varepsilon_{t+1}$ is a zero-mean disturbance term. 
The most commonly followed approaches are the use of individual bivariate regressions using one variable at a time from the Goyal and Welch (GW)
predictor variables \cite{welch2007comprehensive}, or a multivariate regression, which includes the full set of GW predictors in (1) (see \cite{goyal2003predicting}\cite{welch2007comprehensive}\cite{campbell2007predicting} for a bivariate regression and \cite{rapach2010out}\cite{neely2014forecasting}\cite{buncic2017macroeconomic} for a multivariate regression). %Neely et al. (2014) and many others). 

Deep learning models are on the rise, showing impressive results in modeling the complex behavior of financial data. Examples include stock prediction based on long short-term memory (LSTM) networks \cite{fischer2018deep},
deep portfolios based on deep autoencoders \cite{heaton2017deep},
threshold-based approaches using recurrent neural networks \cite{lee2018threshold}, and
deep factor models involving deep feed-forward networks \cite{nakagawa2018deep}, LSTM networks \cite{nakagawa2019deep}, and fundamentals \cite{alberg2017improving}. These studies apply hand-tuned hyper-parameters.

In section \ref{sec:2}, we provide the data used in this study and the preprocessing methods.
In section \ref{sec:3}, we describe the experimental setting and its implementation.
In section \ref{sec:4}, we provide the experimental results and make comparisons between models.
Finally, some concluding remarks are given in section \ref{sec:5}.

\section{Data and preprocessing}
\label{sec:2}
We used sets of fundamentals and technical indicators that have traditionally been used for studying stock predictability. 
\\
\\
\noindent {\bf Technical indicators} 
Technical analysis is a method for forecasting price movements using past prices and volume and includes a variety of forecasting techniques such as a chart
analysis, cycle analysis, and computerized technical trading systems. 

Technical analysis has a long history of widespread use by participants in
speculative markets 
\cite{smidt1965amateur}
\cite{billingsley1996benefits}
\cite{fung1997information}
\cite{menkhoff1997examining}
\cite{cheung2001currency}
\cite{gehring2003technical}, and
there is a large body of academic evidence
demonstrating
%along with critical view \cite{fam70,fam66,van67,van68,jen70},
the usefulness of a technical analysis, including theoretical support 
\cite{brown1989technical} and empirical evidence 
\cite{lo2000foundations}\cite{blume1994market}, as well as their role in out-of-sample equity premium predictability 
\cite{baetje2016equity}
\cite{rapach2010out}
\cite{neely2014forecasting}.

The monthly market data for the S$\&$P500 were obtained from Yahoo Finance and contain daily trading data, i.e., the opening prices, high prices, low prices, adjusted closing prices, and end-of-day volumes. The data are from the period between January 1, 1950 and December 31, 2017 (Fig. \ref{fig:log_return}).
We used a full set of 14 technical indicators based on 3 types of popular technical strategies, moving average crossover rules, momentum rules, and volume rules:
\begin{itemize}
 \setlength\itemsep{1em}
\item	The time-series momentum indicator, MOM($m$), is the generation of a buy signal when the price is higher than the historical price. Its validation is supported by the observation that the ``trend'' effect persists for approximately 1 year and then partially reverses over a longer timeframe.  
Here, $\textrm{MOM}_{t}(m)$ at time $t$ is
defined as follows:
\begin{equation}
 \textrm{MOM}_{t}(m)=\begin{cases}
    1 \textrm{ (Buy signal) }, & \text{if} \quad P_{t} \geq P_{t-m}\\
    -1 \textrm{ (Sell signal) }, & \text{otherwise}.
  \end{cases}
\end{equation}
where $P_{t}$ is the index value at time $t$, and $m$ is the
look-back period. 
We use $m = 1,3,6,9$ and $12$, which are respectively labeled as
$\textrm{MOM}_{t}$(1M), 
$\textrm{MOM}_{t}$(3M),
$\textrm{MOM}_{t}$(6M),
$\textrm{MOM}_{t}$(9M), and
$\textrm{MOM}_{t}$(12M).
	 
\item The moving average indicator, MA$(s,l)$, 
provides a signal for an upward or downward trend.
A buy signal is generated when the short-term moving average crosses above the long-term moving average because this represents the beginning of an upward trend. A sell signal is generated when the short-term moving average 
crosses below the long-term moving average because this represents the beginning of a downward trend.

Let us define
a simple moving average of the index as follows:
\begin{equation}
\textrm{MA}_{j,t}^{P}=(1/j)\sum_{i=0}^{j-1}P_{t-m}  
\textrm{ for } j=s \textrm{ or }l,
\end{equation}	
where $s$ and $l$ are the look-back periods for short and long moving averages. 
%Then, for short 
%$P_{t}$ is the level of the stock price index at time $t$, and
%$s$ ($l$) are the lengths of the short (long) moving averages.
The moving average indicator $\textrm{MA}_{t}(s,l)$
is then designed as follows:
\begin{equation}
 \textrm{MA}_{t}(s,l)=\begin{cases}
    1 \textrm{ (Buy signal) }, & \text{if} \quad \textrm{MA}_{s,t}^{P} \geq \textrm{MA}_{l,t}^{P}\\
    -1 \textrm{ (Sell signal) }, & \text{otherwise}.
  \end{cases}
  \end{equation}
	The six moving average indicators are constructed for $s=1$, $2$, $3$, and 
$l=9$, $12$, which are symbolized as 
	MA(1M-9M), MA(1M-12M), MA(2M-9M), MA(2M-12M), MA(3M-9M), and MA(3M-12M).
\item The volume indicator, VOL($s,l$), indicates a strong market
trend if the recent stock market volume and stock price increase.
Let us define the on-balance volume (OBV) as follows:
\begin{equation}
\textrm{OBV}_{t}=\sum_{k=1}^{t}VOL_{k}D_{k},
\end{equation}
where $VOL_{k}$ is a measure of the trading volume (i.e., number of shares traded) during period $k$, and $D_{k}$ is a binary variable: 
\begin{equation}
  D_{k}=\begin{cases}
    1, & \text{if} \quad P_{k}\geq P_{k-1}\\
    -1, & \text{otherwise}.
  \end{cases}
\end{equation}
The value of $\textrm{OBV}_{t}$ conceptionally measures both positive and negative 
volume based on the belief that changes in volume can predict a stock movement. The volume-based indicator is then defined as the difference between
the moving averages with a $s$-period and $l$-period:
\begin{equation}
 \textrm{VOL}(s,l)=\begin{cases}
    1 \textrm{ (Buy signal) }, & \text{if} \quad \textrm{MA}_{s,t}^{\textrm{OBV}} \geq \textrm{MA}_{l,t}^{\textrm{OBV}}\\
    -1 \textrm{ (Sell signal) }, & \text{otherwise}.
  \end{cases}
\end{equation}
Here,
$
\textrm{MA}_{j,t}^{\textrm{OBV}}=(1/j)\sum_{i=0}^{j-1}\textrm{OBV}_{t-i}
$ is the moving average of $\textrm{OBV}_{t}$ for $j=s$ or $l$. 
The six moving average indicators are constructed for $s=1$, $2$, $3$ and 
$l=9$, $12$, which are symbolized as 
VOL(1M-9M), VOL(1M-12M),
VOL(2M-9M), VOL(1M-12M), VOL(3M-9M) and VOL(3M-12M).
\end{itemize}
\noindent {\bf Fundamental indicators}
We use the financial indicators employed by 
\cite{welch2007comprehensive} for the
U.S. stock market, which is available from Amit Goyal's web site. 
We use updated data consisting of 14 popular fundamental variables
spanning from January 1950 to December 2017. We provide a short definition of
these variables as follows.
\begin{itemize}
 \setlength\itemsep{1em}
\item Dividend-price ratio, DP: Log of a 12-month moving sum of dividends paid on the S\&P 500 index minus the log of the stock prices. 
\item Dividend yield, DY: Log of a 12-month moving sum of dividends minus the log of 1-month lagging stock prices.
\item Earning-price ratio, EP: Log of a 12-month moving sum of earnings on the S\&P 500 index minus the log of the stock prices.
\item Dividend-payout ratio, DE: Log of a 12-month moving sum of dividends minus the log of a 12-month moving sum of earnings.
\item Stock variance, SVAR: Sum of squared daily returns on the S$\&$P500. 
\item Book-to-market ratio, BM: Ratio of book value to market value for the Dow Jones Industrial Average. 
\item Net equity expansion, NTIS: Ratio of 12-month moving sum of net issues by NYSE listed stocks divided by their total market capitalization.
\item Treasury Bill rate, TBL: Interest rate on a 3-month treasury bill from the secondary market. 
\item Long-term yield, LTY: Long-term government bond yields. 
\item Long-term rate of return, LTR: Long-term government bond returns 
\item Term spread, TMS: Difference between the long and term yield on government bonds and T-bills. 
\item Default yield spread, DFY: Difference between BAA- and AAA-rated corporate bonds and returns on long-term government bonds. 
\item Default return spread, DFR: Difference between the return on long-term corporate bonds and returns on the long-term government bonds.
\item Inflation, INFL: Consumer Price Index (CPI) for all urban consumers.\\ 
\end{itemize} 
\section{Experiments}
\label{sec:3}
\noindent {\bf Data Splits:}
As mentioned earlier, the predictability found in traditional studies is not uniform over time and is concentrated within certain periods \cite{neely2014forecasting}. To check the robustness, we investigated the predictability over four different periods, the entire period of $1950-2017$ (Exp. 1) and its sub-periods of $1950-2015$ (Exp. 2), $1950-2007$ (Exp. 3), and $1950-2002$ (Exp. 4).
For each experiment, we split the
data into in-sample and out-of-sample periods.
The in-sample data were divided into a training dataset (50$\%$) for developing the prediction models and a validation set (50$\%$) for evaluating its predictive ability.
\\
 \\
\noindent{\bf Training:} 
Deep feedforward neural networks (DNNs) were used in this study. We applied TPE for automated hyperparameter tuning with
additional tests using simulated annealing and a random search to further confirm our results. 
The hyperparameters and their
prior distributions are summarized in Table \ref{params}.
For hyperparameter selection, we trained DNNs on an in-sample training set and selected the model with the lowest validation error.
We limited the number of function evaluations for finding optimal hyper-parameters to $50$. 
Each evaluation comprised training the DNN
models for 200 epochs and selecting the model with the lowest validation error.
\\
\\
\noindent{\bf Regularizer:} 
We are particularly interested in regularization methods for model generalization
because the time-dependent behavior of financial data is likely to cause a parameter instability over an out-of-sample. 
We examined the effectiveness of the most popular regularization methods, namely, a dropout and batch normalization (BN). 
A dropout\cite{srivastava2014dropout} 
is a simple way to prevent co-adaptation among
hidden nodes of deep feed-forward neural networks by randomly dropping out selected hidden nodes.
In recent years, batch normalization \cite{IoffeS2015batch} has replaced a dropout
in modern neural network architectures. 
It uses the distribution of
the summed input to a neuron over a mini-batch of training cases to compute the
mean and variance, which are then used to normalize the summed input to the
neuron for each training case.
Dropout and BN layers are employed for all hidden layers.
\\
\\
\begin{table}[htbp]
\centering
\caption{List of parameters and their corresponding range of
values used in the grid search.}
\label{table:meanerrorbaseline}
\small
\begin{tabular}{lll}
\toprule
  Hyperparamter &\quad \quad \quad & Considered values/functions \\
\midrule
    Number of Hidden Layers && \{2, 3\} \\
    Number of Hidden Units && \{2, 4, 8, 16\} \\
     \makecell[l]{ Standard deviation} &&\{0.025,0.05,0.075\}\\
    Dropout  && \{0.25, 0.5, 0.75\} \\
    Batch Size && \{28, 64, 128\} \\
    Optimizer && \{RMSProp, ADAM, SGD (no momentum)\} \\
    Activation Function&& Hidden layer: \{tanh, ReLU, sigmoid\}, Output layer: Linear \\
    Learning Rate && \{0.001\} \\
    Number of Epochs && \{100\} \\ 
\bottomrule
\end{tabular}
\parbox{\textwidth}{\small%
\vspace{1eX} % If wanted space after the bottomrule
{\bf Number of Layers}: number of the layers of the neural network.
{\bf Number of Hidden Units}: number of units in the hidden layers
of the neural network.
{\bf Standard Deviation}: standard deviation of a random normal initializer. 
{\bf Dropout}: dropout rates. 
{\bf Batch Size}: number of samples per 
batch.  
{\bf Activation}: sigmoid function $\sigma(z)=1/(1+e^{-z})$, hyperbolic 
tangent function $\textrm{tanh}(z)=(e^{z}-e^{-z})/(e^{z}-e^{-z})$,
and rectified linear unit (ReLU) function $\textrm{ReLU}(z)=\textrm{max}(0,z)$.
{\bf Learning Rate}: learning rate of the back-propagation algorithm.
{\bf Number of Epochs}: number of iterations for all of the training data.
{\bf Optimizer}: stochastic gradient descent (SGD) \cite{kingma2014adam}, RMSProp \cite{tieleman2012lecture}, and ADAM \cite{kingma2014adam}}
 \label{params}%
\end{table}

\noindent {\bf Out-of-sample $R^{2}$ statistic:} 
We measured the out-of-sample $R^{2}$ statistics ($R_{\textrm{OS}}$) \cite{campbell2007predicting} for a comparison with the in-sample $R^{2}$ statistics ($R^{2}_{\textrm{IS}}$)
and evaluated the forecasting power of the models. 
The $R_{\textrm{OS}}^{2}$ statistic measures the improvement in the mean square
forecast error (MSFE) for the return forecast relative to the simple historical
average (or constant expected return) forecast, which ignores information contained in the predictors. This is computed as follows:
\begin{equation}
R^{2}_{\textrm{OS}}=1-\frac{\sum_{t=1}^{T}(r_{t}-\hat{r}_{t})^{2}}{\sum_{t=1}^{T}(r_{t}-\bar{r}_{t})^{2}},
\end{equation} 
where $\hat{r}_{t}$ is the fitted value from a predictive regression estimated through period $t-1$, and $\bar{r}_{t}$ is the historical average return estimated through period $t-1$.
\\
\\
\noindent {\bf Model stability:} We analyzed the model stability over time in terms of the feature importance. 
Stock price dynamics is so complex with complicated interactions among changing micro
behavior, varying product cycles, interdependent industrial structures, and cyclic macro environment, thus it leads to gradual or sudden shifts in the model parameters. For example, traditional univariate models are highly exposed to
the model instability in the in-sample, which demonstrates the time-dependency of 
the statistical significance and the coefficient of the predictor variables \cite{neely2014forecasting}. To overcome this problem, a multivariate regression model is proposed through which the changes to the parameters at breaks are estimated \cite{paye2006instability}.

We examined the stability of the trained model over time by 
computing the SHapley Additive exPlanation (SHAP) values of the features  \cite{lundberg2018consistent}
to find the contribution of the features in the prediction and determine the change in ranking of the features over time. 
\section{Results}
\label{sec:4}

\subsection{Technical Indicators}

\subsubsection{Dropout versus batch normalization}
\begin{figure}[t]
\centering
     \scalebox{0.5}
  {
	\includegraphics{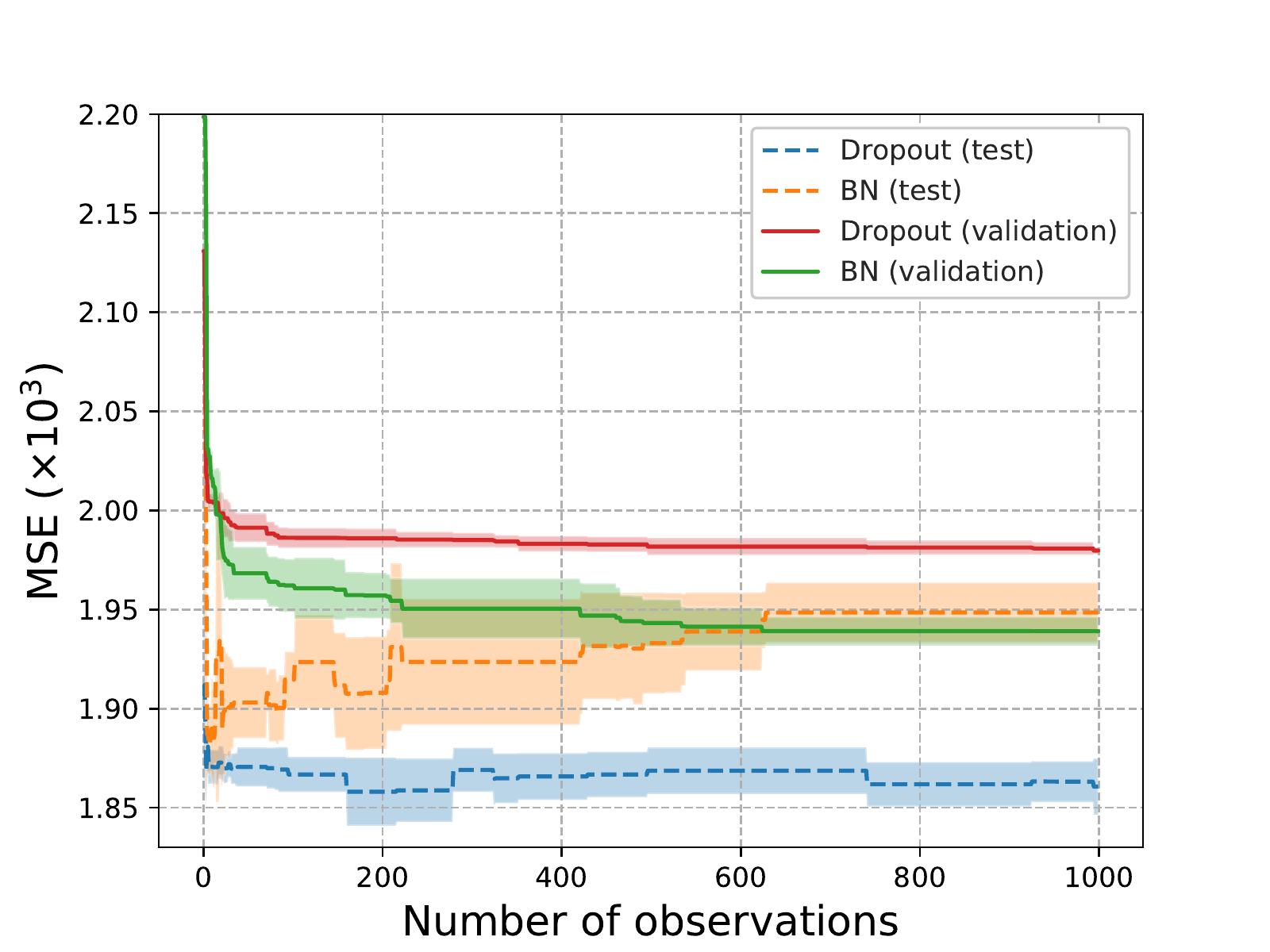}

   }
\caption{Validation and testing errors of the DNNs with dropout and with BN with regards to $50$ function evaluations and 200 epochs for each function evaluation. 
The (dashed) lines are the average score over
five random initializations and 
the shaded regions correspond to one standard deviation.
\iffalse
The (dashed) lines represent the
average over five random initializations.
\fi
\iffalse
of five repetitions with different training and validation splits, and the shaded
areas represent the standard deviation over those repetitions.
The results are the average testing score
over five trials where the shaded regions correspond to the
standard deviation.
\fi}
\label{MSE_TPE_BN}
\end{figure}

\begin{figure}[t]
\centering
     \scalebox{0.5}
  {
	\includegraphics{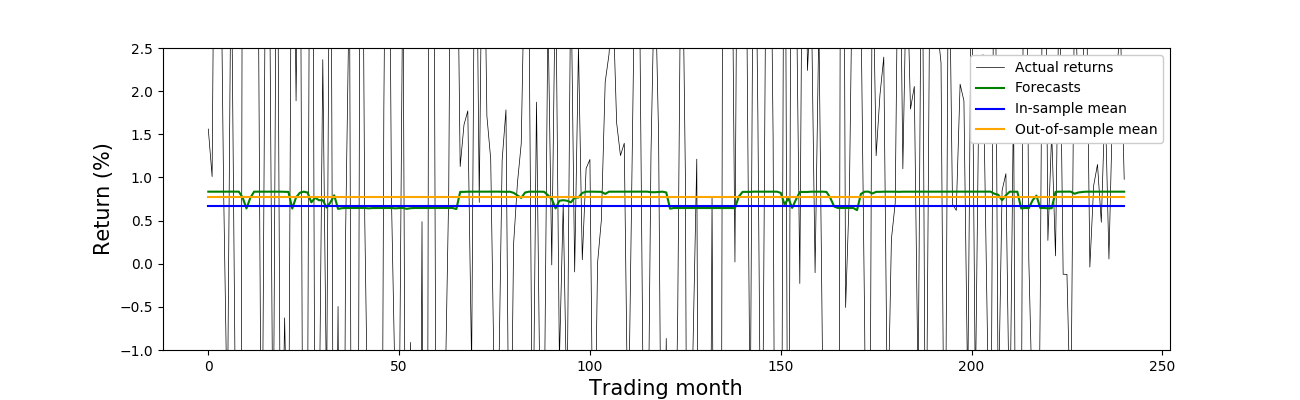}
   }
\caption{Comparison of actual and predicted values over the out-of-sample period. The actual return is drawn by the thin solid black line.
Forecasted values from the DNN with dropout,
in-sample mean, and
out-of-sample mean 
are drawn by the solid green, blue and yellow lines, respectively.
}
\label{MSE_forecast_pattern}
\end{figure}

We compared a DNN with a dropout and a DNN with batch normalization for the four experiments. The following observations can be made regarding the results reported in Table \ref{tab:dropout}.
\begin{itemize}
\item Both DNNs show a good in-sample predictive power of a positive $R^{2}_{\textrm{IS}}$ for all experiments. The in-sample predictive power of the BN ranging over 1.740 to 2.968 is stronger than that of the dropout ranging over 0.424 to 0.748. 
\item The DNN with a dropout achieves a good out-of-sample predictive power, showing positive $R^{2}_{OS}$ values for all experiments, which means that it outperforms the historical mean return over the training and validation periods.
However, the BM model achieves a poor out-of-sample predictive power, with negative $R^{2}_{OS}$ values for all experiments. A dropout is more effective at preventing a model instability.
\item The instability of the BN model is derived from an overfitting to the in-sample set based on the observation that, although $\textrm{MSE}_{\textrm{train}}$ and $\textrm{MSE}_{\textrm{val}}$ of the BN model are lower than those of the dropout model (except for only $\textrm{MSE}_{\textrm{train}}$ in Exp. 2), $\textrm{MSE}_{\textrm{test}}$ of the BN model is higher than that of the dropout model. Figure \ref{MSE_TPE_BN} graphically shows the overfitting occurring during the training in Exp. 1.
\item The results indicate that an in-sample predictive content does not necessarily translate into an out-of-sample predictive ability, nor ensure the stability of the predictive relation over time.
\item The degree of predictability varies according to the experimental period, showing that Exp. 2 and 3 show a strong predictability of $1.889$ and $1.670$, and Exp. 1 and 4 show a relatively weak predictability of $0.569$ and $0.319$, respectively.
\item Figure \ref{MSE_forecast_pattern} graphically shows how to beat the historical average in Exp. 1. The dropout model forecasts returns around the mean of the out-of-sample, whereas the historical average showed a greater deviation. This means the model can be adjusted better to a new market environment than the historical average. 
\item The DNN with a dropout achieves an average predictability of 0.53$\%$ in-sample and 1.11$\%$ out-of-sample. The DNN with a dropout has an average predictability of 2.312$\%$ in-sample 
and $-2.8545\%$ out-of-sample.
\end{itemize}

\begin{table}[htbp]
\small
  \centering
  \caption{Comparison of models based on average prediction performance ($\pm$1 s.d. in parentheses) over 5 runnings with different random initial seeds for each experiment.}
       \begin{tabular}{L{3.cm}  C{2.cm} C{2.cm} C{2.cm} C{2.cm} C{2.cm} }
            \toprule
     Model  & \multicolumn{1}{c}{$\textrm{MSE}_{\textrm{train}}$  }    & \multicolumn{1}{c}{$\textrm{MSE}_{\textrm{val}}$  } & \multicolumn{1}{c}{$\textrm{MSE}_{\textrm{test}}$ } &
      \multicolumn{1}{c}{$R^{2}_{IS}$ } & \multicolumn{1}{c}{$R^{2}_{OS}$ } \\
         \hline
            & \multicolumn{5}{c}{Exp. 1}    \\
            \cline{2-6}
      DNN w. dropout & \makecell{0.129 \\($\pm 3.236 $)}& \makecell{0.197 \\ ($\pm 0.171 $)}  & \makecell{ $\bm{0.186}$ \\ ($\pm\bm{ 1.506 }$)}& \makecell{0.748 \\ ($\pm$1.040)}& \makecell{$\bm{0.569}$ \\($\bm{\pm 0.621}$)} \\
         DNN w. BN & \makecell{ $\bm{0.128}$\\($\bm{\pm 0.646}$)}& \makecell{$\bm{0.193}$\\($\bm{\pm 1.333 }$)} & \makecell{0.194\\($\pm 1.713$)} & \makecell{$\bm{1.740}$ \\($\bm{\pm 0.247$})} &\makecell{$-3.74$ \\($\pm$0.804)} \\
         \hline
            & \multicolumn{5}{c}{Exp. 2}    \\
            \cline{2-6}
      DNN w. dropout & \makecell{ $\bm{0.126}$ \\ ($\bm{\pm 0.062}$)} & \makecell{ 0.206 \\ ($\pm 0.130$)}  &\makecell{$\bm{0.201}$ \\ ($\bm{\pm 0.540}$)}& \makecell{0.451 \\($\pm$0.028)}& \makecell{$\bm{1.889}$ \\ ($\bm{\pm 0.242)}$} \\
         DNN w. BN & \makecell{0.127\\($\pm 1.739$)}&  \makecell{ $\bm{0.201}$\\($\bm{ \pm 0.298}$)} &\makecell{0.209\\($\pm 3.240$)} & \makecell{$\bm{1.890}$\\ ($\bm{\pm 0.293}$)} &\makecell{$ -2.70$\\ ($\pm$0.906)} \\
         \hline
            & \multicolumn{5}{c}{Exp. 3 }    \\
            \cline{2-6}
        DNN w. dropout & \makecell{0.130\\($\pm 0.189 $)} & \makecell{0.216 \\($\pm 0.038 $)}  & \makecell{$\bm{0.147}$ \\($\bm{\pm 0.174}$)}&\makecell{0.507 \\($\pm$0.056)} & \makecell{$\bm{1.670}$ \\($\bm{\pm 0.143}$)} \\
         DNN w. BN & \makecell{$\bm{0.125}$\\($\bm{\pm 1.618 }$)}&  \makecell{$\bm{0.213}$\\($\bm{\pm 0.768 }$)} & \makecell{0.153\\($\pm 2.781 $)} &\makecell{$\bm{2.650}$ \\($\bm{\pm 0.543}$)} & \makecell{$-3.518$ \\($\pm$2.650)} \\
         \hline
                     & \multicolumn{5}{c}{Exp. 4}    \\
                     \cline{2-6}
        DNN w. dropout & \makecell{0.1193 \\($\pm 0.697$)}& \makecell{0.197 \\($\pm 0.716$)} & \makecell{$\bm{0.218}$ \\($\bm{\pm 0.356}$) }& \makecell{0.424 \\($\pm$0.181)} & \makecell{$\bm{0.319}$ \\($\bm{\pm 0.139}$)}\\
        DNN w. BN & \makecell{$\bm{0.115}$\\($\bm{\pm 3.998 }$)}&  \makecell{$\bm{0.196}$\\($\bm{\pm 0.702}$)} & \makecell{0.219\\($\pm 2.160$)} & \makecell{$\bm{2.968}$ \\($\bm{\pm 0.959}$)} & \makecell{$-1.460$ \\($\pm$0.989)}\\
            \bottomrule
    \end{tabular}%
\begin{tablenotes}[flushleft]\footnotesize
    \item[]
    Note: All the $\textrm{MSE}$ and $R^{2}$ values have been multiplied by a factor of $10^{-2}$ and all the s.d. values has been multiplied by a factor of $10^{-5}$.
    \end{tablenotes}
  \label{tab:dropout}%
  \end{table}%

\subsubsection{Effect of optimizer choice}
\begin{figure}[t]
\centering
        \scalebox{0.55}
  {
	\includegraphics{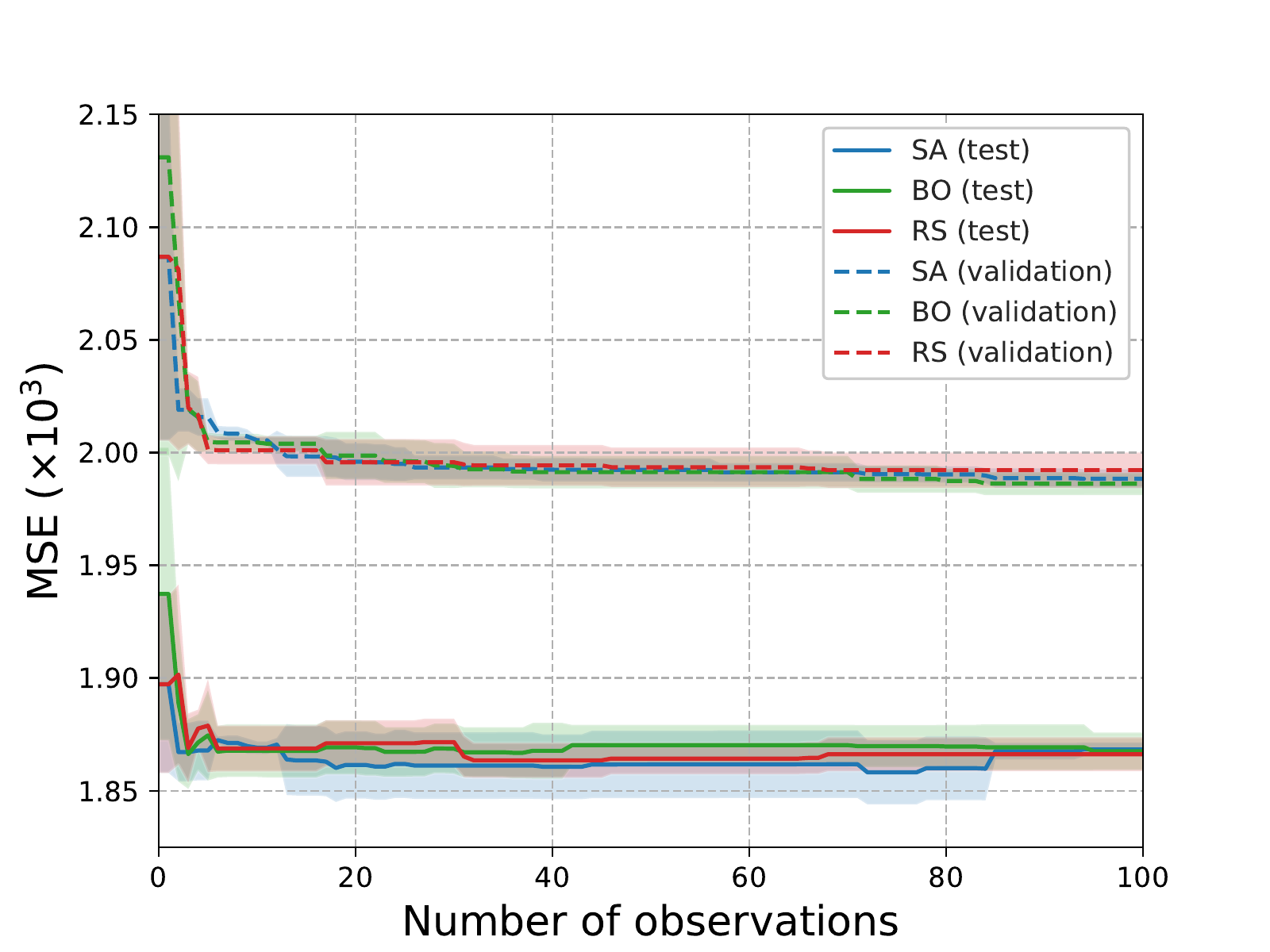}
   }
     
\caption{Comparison of the simulated annealing (SA), TPE, and random search (RS) performances on the validation and test sets for the first 100 observations. 
The (dashed) lines are the average score over
five random initializations and 
the shaded regions correspond to one standard deviation.
}
\label{MSE_three_opt}
\end{figure}

To further check the robustness of a dropout with respect to the dependency on the selected optimization algorithm, we repeated the experiments using a random search and simulated annealing. As shown in
Fig. \ref{MSE_three_opt}, a comparable performance is shown for both the validation and test set, without an overfitting to the former.
Our observations on different optimizers consistently suggest that a dropout helps improve the generalization. This indicates that the benefits of the HPO are general, without
depending on a specific optimizer, thereby demonstrating its robustness.

\subsubsection{Model stability over time}
\begin{figure}[h]
   \advance\leftskip-1cm
  \centering
  \subfigure[DNN with dropout. From left, Exp. 1–4.]{%
    \includegraphics[width=0.25\linewidth]{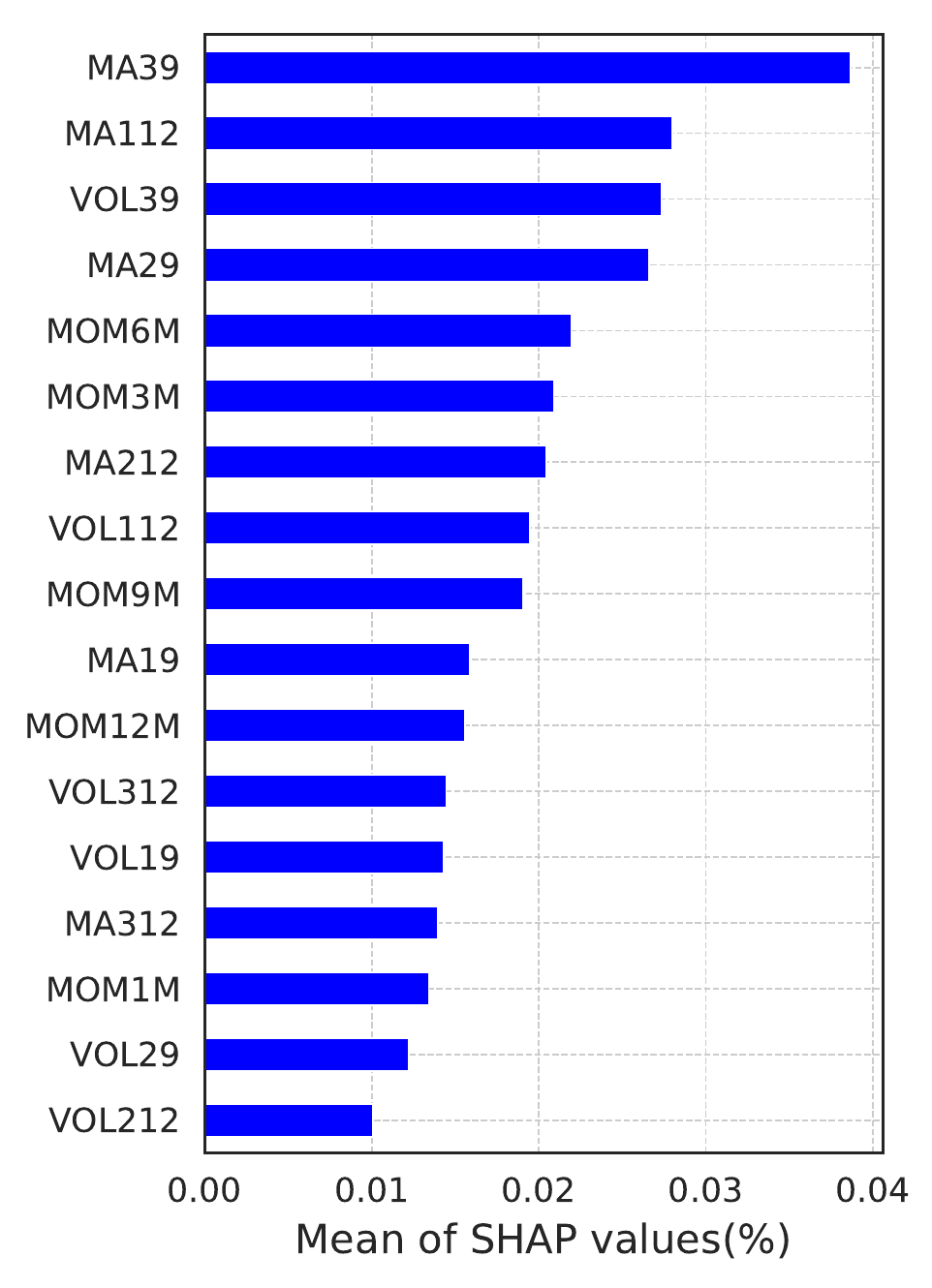}%
    \includegraphics[width=0.25\linewidth]{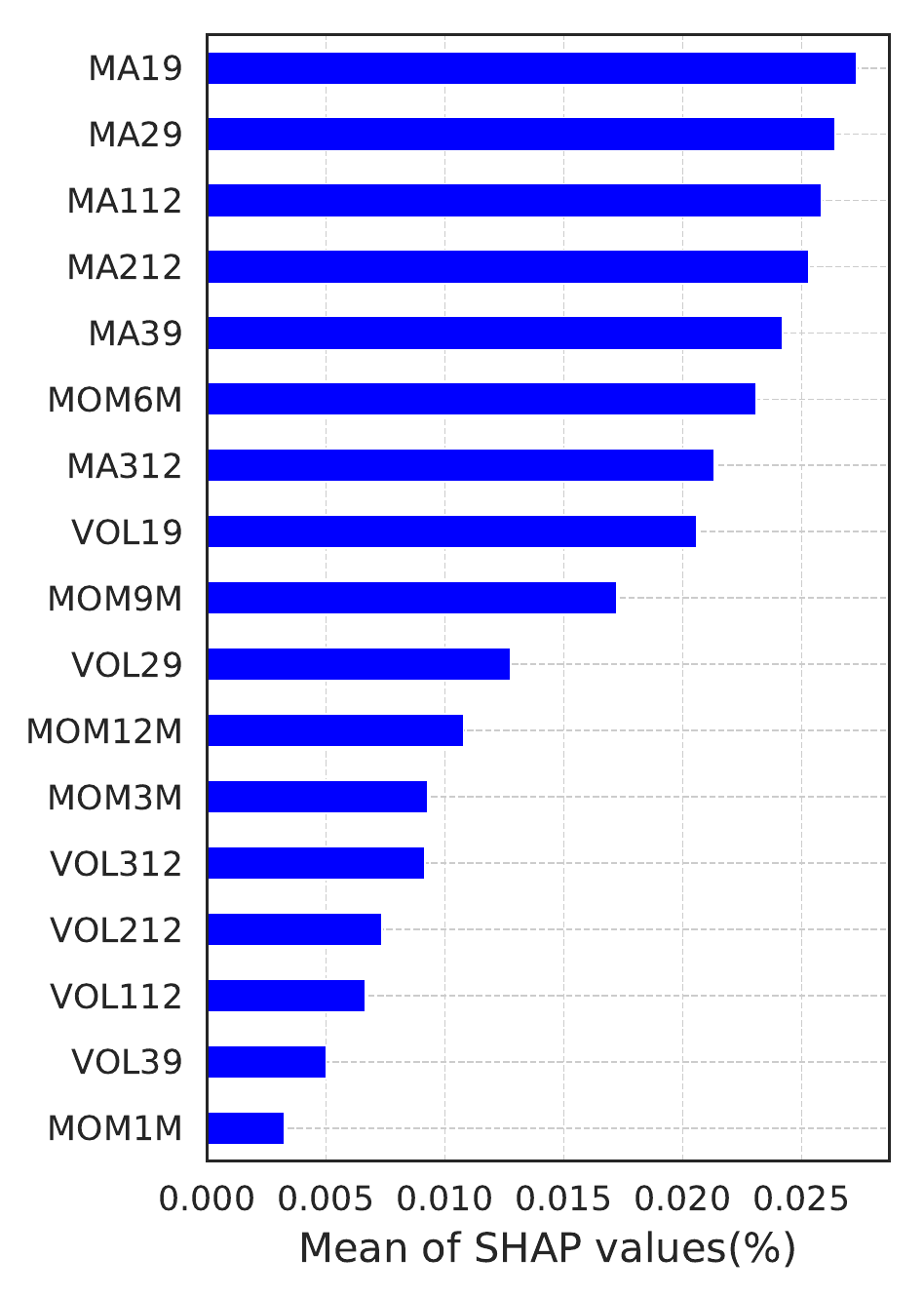}%
    \includegraphics[width=0.25\linewidth]{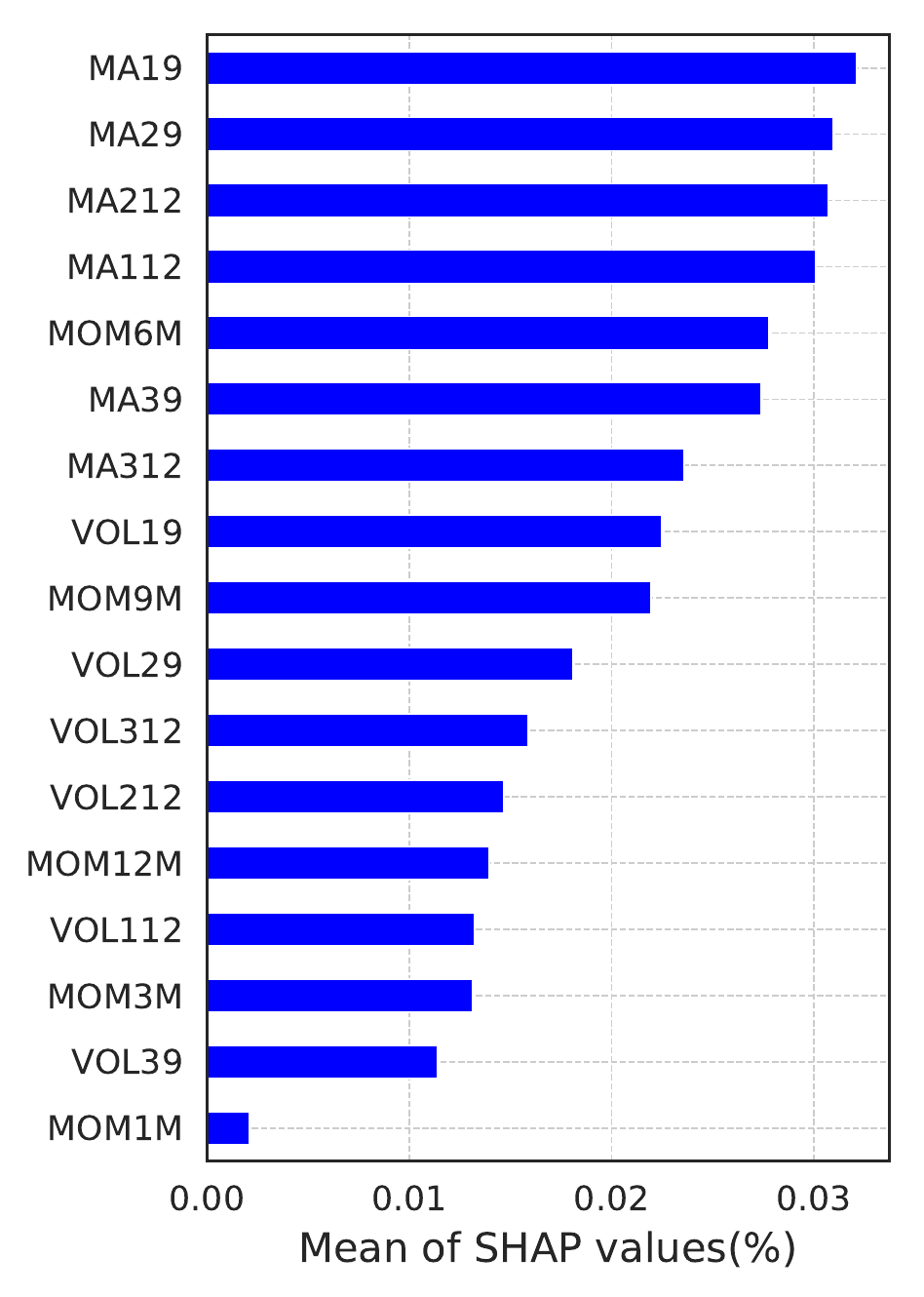}%
    \includegraphics[width=0.25\linewidth]{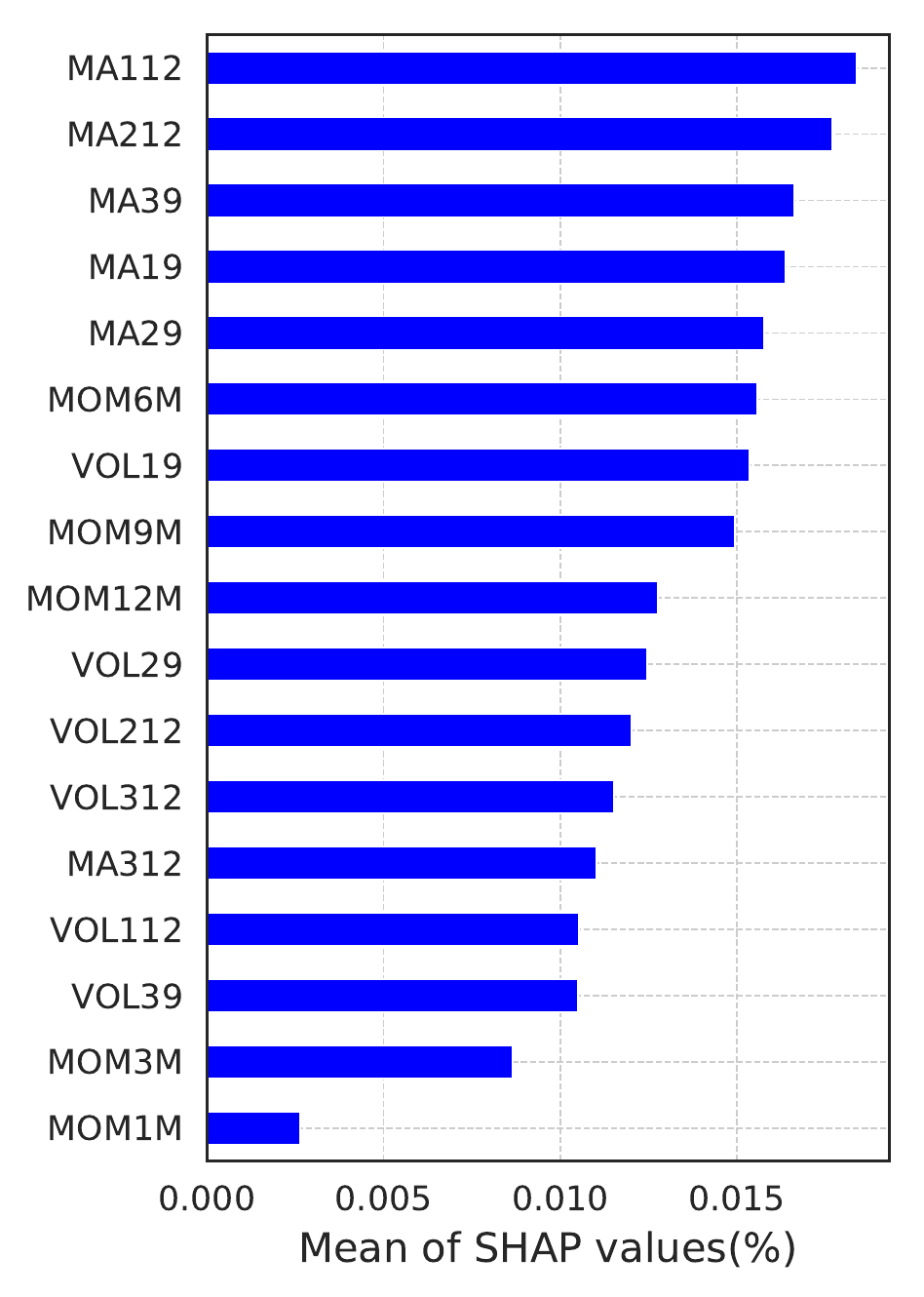}%
  }\\
  \subfigure[DNN with BN. From left, Exp. 1–4.]{%
    \includegraphics[width=0.25\linewidth]{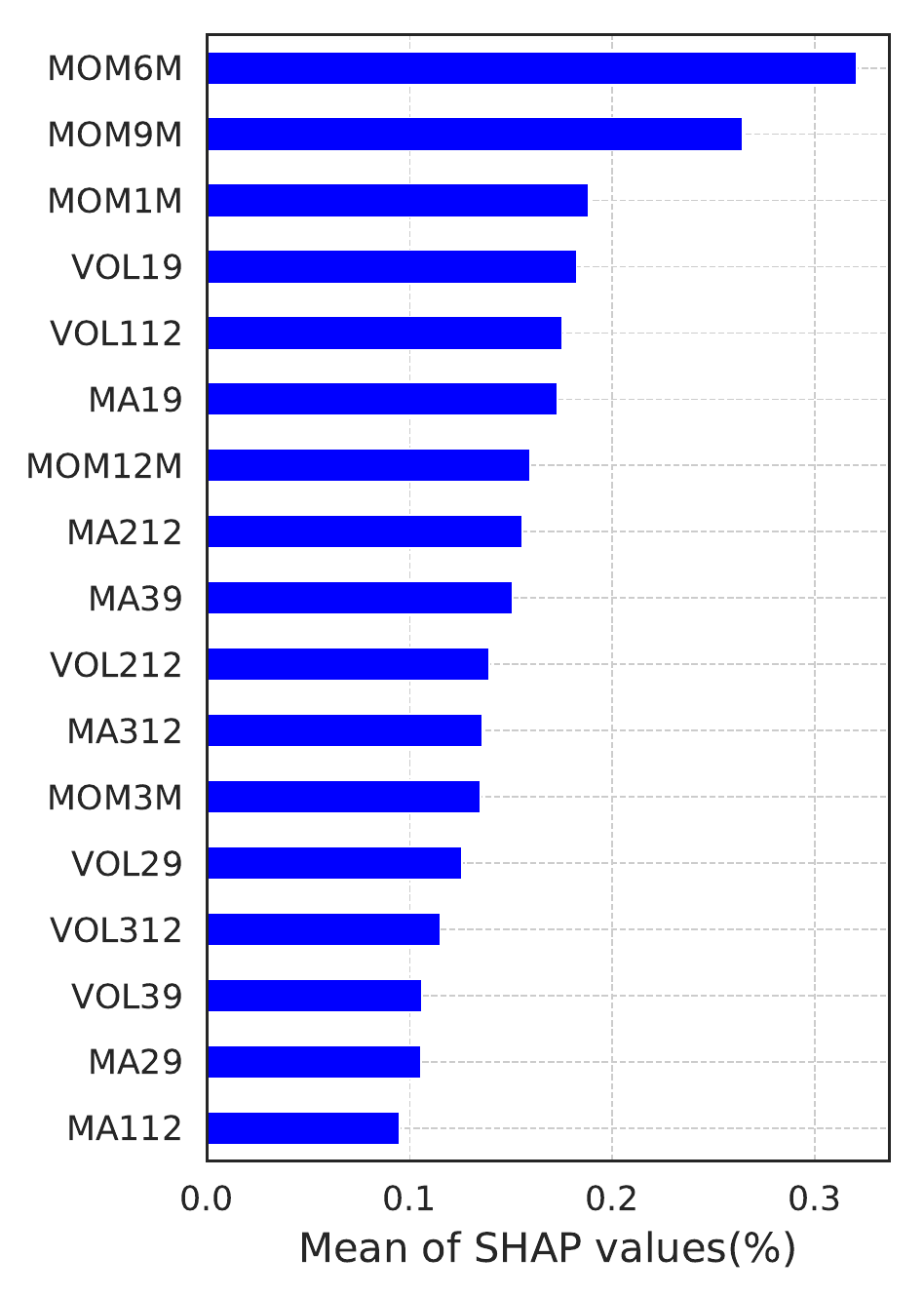}
    \includegraphics[width=0.25\linewidth]{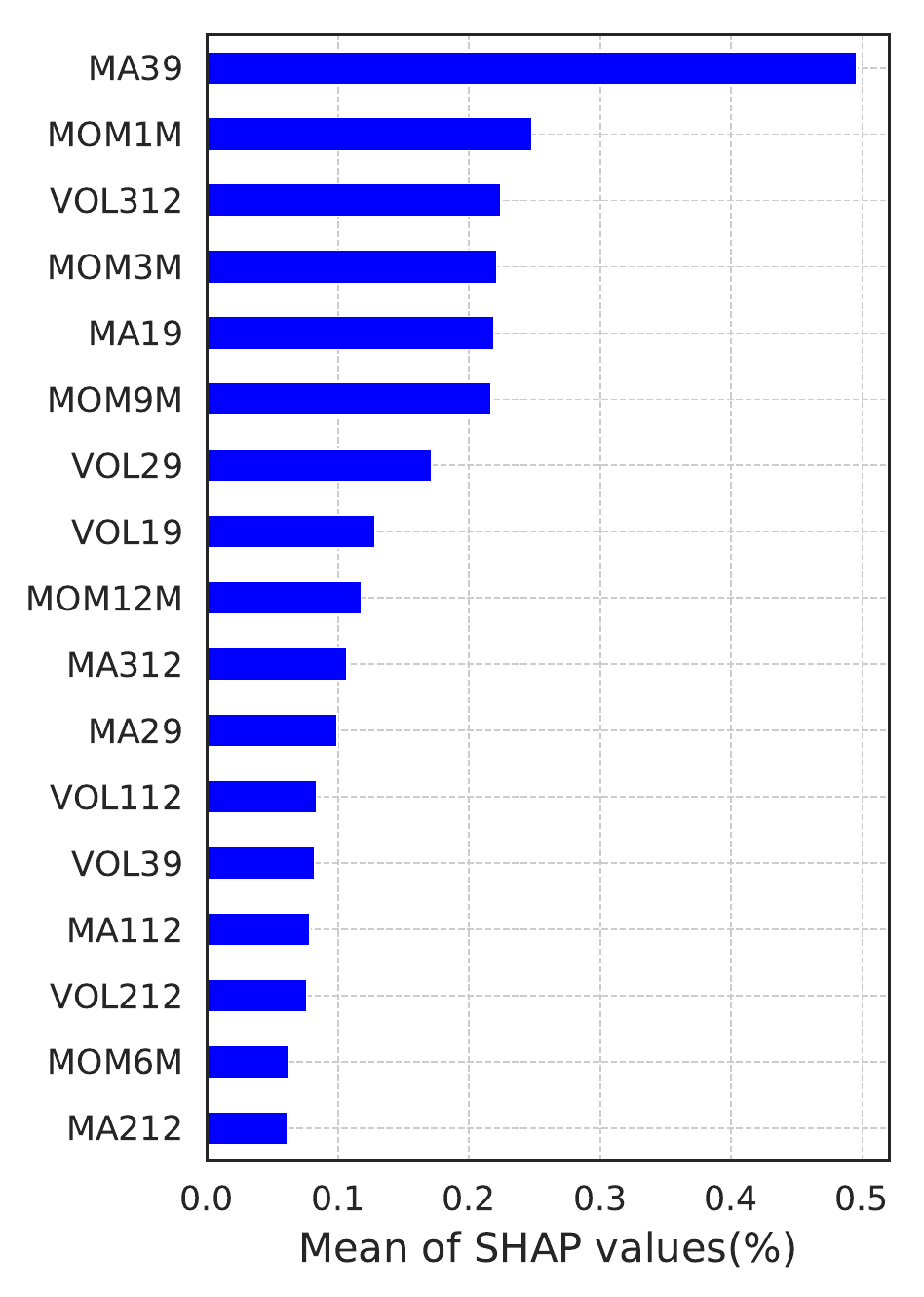}
    \includegraphics[width=0.25\linewidth]{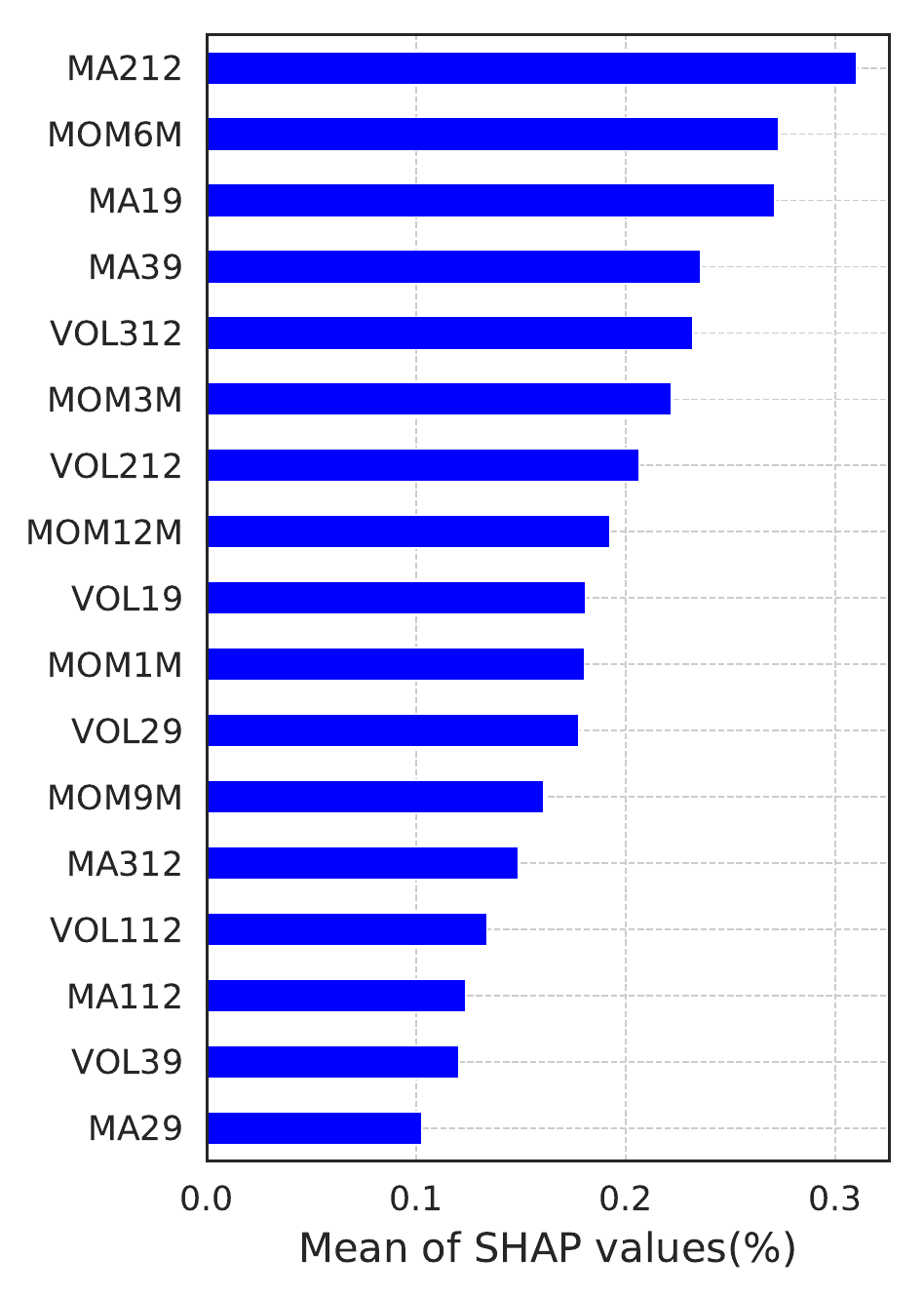}
    \includegraphics[width=0.25\linewidth]{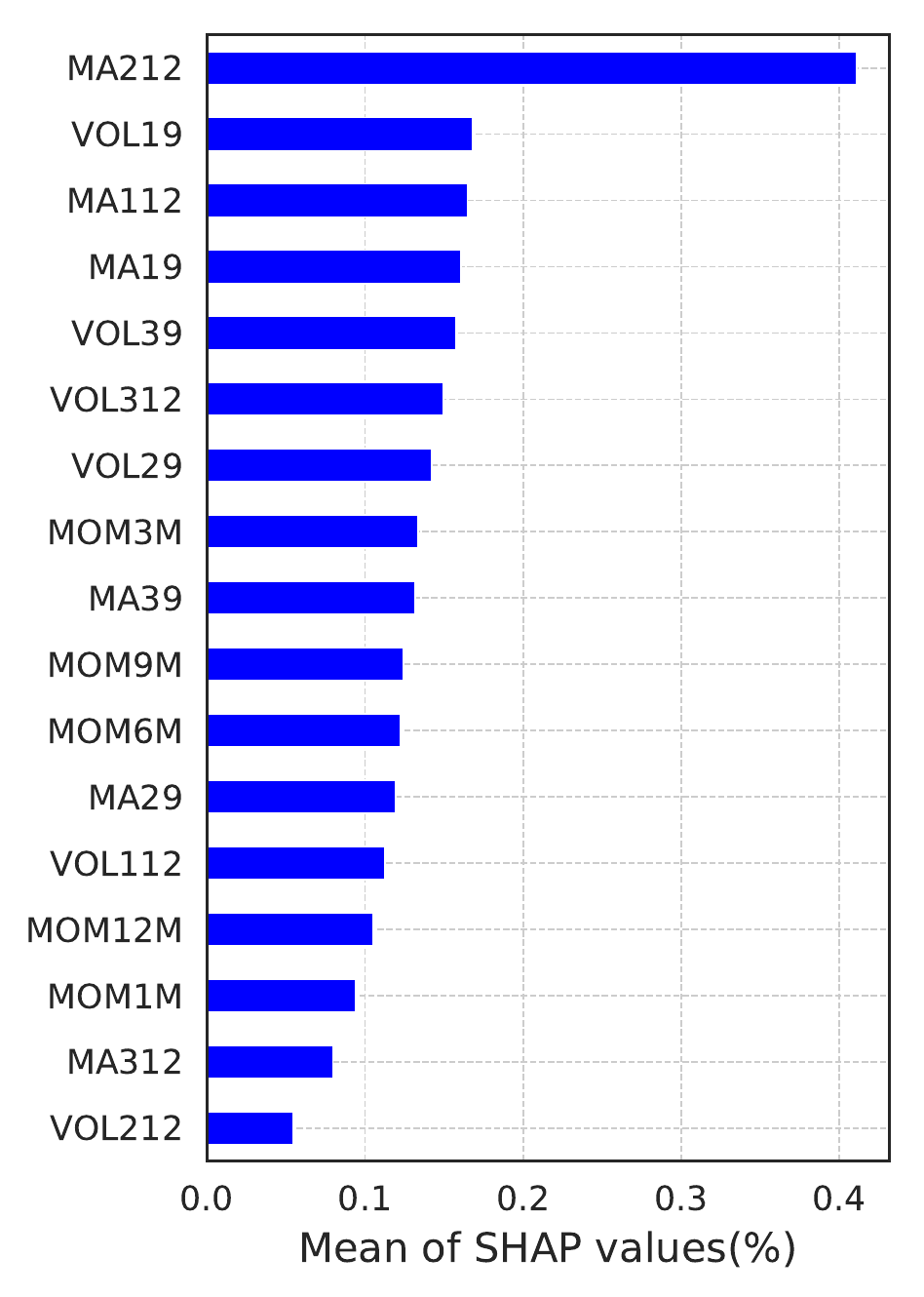}
    }
      \caption{Mean absolute value of SHAP values for each features for Exp. 1--4.}
  \label{fig_SHAP}
\end{figure}

Figure \ref{fig_SHAP} shows the importance of features arranged in decreasing order for the dropout and batch normalization models. They were calculated by summing the absolute values of the SHAP values. Table \ref{SHAP_rank} shows the rank of the features over time from Exp. 4 to Exp. 1. The following observation was made based on the results.
 \begin{itemize}
 \item The feature importance is sensitive to the selected experimental periods for both models. This implies that the selection of a small number of features based on their importance can prevent a model generalization for unseen (new) data. 
  
\item Overall, we observed that a DNN with a BN 
achieves a greater variability than a DNN with a dropout. In the experiments with a dropout, the five variables $\{$MA112, MA212, MA39, MA29, MOM6M$\}$ and
the six variables $\{$MOM12M, VOL29,  VOL212, VOL312, MOM3M, MOM1M $\}$
remain in the top half (from 1st to 8th) and bottom half across the experiments, respectively.
By contrast, in the experiments using a BN, only $\{$MA19$\}$ remains in the top half and $\{$MA29, MA312$\}$ remain in the bottom half. This indicates that a DNN with a dropout is more generalized against a time change and explains the outperformance of $R^{2}_{OS}$ in a more fundamental manner.
 \end{itemize}

 \begin{table}[!htb]
  \caption{\label{SHAP_rank}
Feature ranking results of DNNs with dropout (left) and with BN (right).}
  \small
    \begin{minipage}{.55\textwidth}
      \centering
 \begin{tabular}{lccccc}
 \toprule
          & \multicolumn{1}{c}{Exp. 4} & \multicolumn{1}{c}{Exp. 3} & \multicolumn{1}{c}{Exp. 2} & \multicolumn{1}{c}{Exp. 1} \\
          \hline
    MA112 & 1     & 4     & 3     & 2     \\
    MA212 & 2     & 3     & 4     & 7    \\
    MA39  & 3     & 6     & 5     & 1     \\
    MA19  & 4     & 1     & 1     & 10     \\
    MA29 & 5     & 2      & 2     & 4     \\
    MOM6M & 6     & 5     & 6     & 5     \\
    VOL19 & 7     & 8     & 8     & 13    \\
    MOM9M & 8     & 9     & 9     & 9     \\
    MOM12M & 9     & 13   & 11    & 11     \\
    VOL29 & 10    & 10    & 10    & 16    \\
    VOL212& 11    & 12    & 14    & 17     \\
    VOL312 & 12    & 11   & 13    & 12     \\
    MA312 & 13    & 7     & 7     & 14    \\
    VOL112& 14    & 14    & 15    & 8     \\
    VOL39 & 15    & 16    & 16    & 3     \\
    MOM3M & 16    & 15    & 12    & 16    \\
    MOM1M & 17    & 17    & 17    & 15     \\
    \bottomrule
    \end{tabular}%
    \end{minipage}
    \begin{minipage}{.3\textwidth}
      \centering
         \begin{tabular}{lcccc}
          \toprule
          & \multicolumn{1}{c}{Exp. 4} & \multicolumn{1}{c}{Exp. 3} & \multicolumn{1}{c}{Exp. 2} & \multicolumn{1}{c}{Exp. 1} \\
          \hline
    MA212 & 1     & 1     & 17    & 8     \\
    VOL19  & 2     & 9     & 8     & 4   \\
    MA112 & 3     & 15    & 14    & 17    \\
    MA19  & 4     & 3     & 5     & 6      \\
    VOL39 & 5     & 16    & 13    & 15   \\
    VOL312& 6     & 5     & 3     & 14    \\
    VOL29 & 7     & 11    & 7     & 13   \\
    MOM3M & 8     & 6     & 4     & 12    \\
    MA39 & 9     & 4    & 1    & 9    \\
    MOM9M & 10    & 12    & 6     & 2      \\
    MOM6M & 11    & 2    & 16    & 1   \\
    MA29 & 12    & 17    & 11    & 16    \\
    VOL112& 13    & 14    & 12    & 5    \\
    MOM12M& 14    & 8    & 9    & 7   \\
    MOM1M & 15    & 10    & 2    & 3    \\
    MA312 & 16    & 13    & 10    & 11    \\
    VOL212& 17    & 7    & 15    & 10     \\
        \bottomrule
    \end{tabular}%
    \end{minipage}
    \iffalse
    \begin{tablenotes}[flushleft]\footnotesize
    \item[]
    Note: Superscripts $t$ and $b$ denote the features remaining
    in the top and bottom halves of the features during all experiments.
    \end{tablenotes}
    \fi
  \end{table}

\subsection{Fundamentals}
\subsubsection{Predictability and model stability}
Table \ref{tab:fundamentals} shows the results produced through the
same procedure as used in the previous experiments applying fundamentals. 
The following observations can be made regarding the results.
\begin{itemize}
\item For both models, fundamental data are prone to an overfitting to the in-sample data as shown in
the positive $R_{IS}^{2}$ and negative $R_{OS}^{2}$ values. 
%The in-sample predictability is better than those of Fundamentals. 
\item A DNN with a dropout outperforms a DNN with a BN in terms of better values of $R_{IS}^{2}$ and $R_{OS}^{2}$ except for only $R_{IS}^{2}$ in Exp. 1.  
\end{itemize}

\begin{table}[htbp]
\small
  \centering
  \caption{Comparison of models based on average prediction performance ($\pm$1 s.d. in parentheses) over 5 runnings with different random initial seeds for each experiment.}
       \begin{tabular}{L{3.cm}  C{2.cm} C{2.cm} C{2.cm} C{2.cm} C{2.cm} }
            \toprule
     Regularizer  & \multicolumn{1}{c}{$\textrm{MSE}_{\textrm{train}}$ }    & \multicolumn{1}{c}{$\textrm{MSE}_{\textrm{val}}$  } & \multicolumn{1}{c}{$\textrm{MSE}_{\textrm{test}}$  } &
      \multicolumn{1}{c}{$R^{2}_{IS}$ } & \multicolumn{1}{c}{$R^{2}_{OS}$ } \\
         \hline
            & \multicolumn{5}{c}{Exp. 1}    \\
            \cline{2-6}
      DNN w. dropout & \makecell{0.129 \\($\pm 0.580 $)}& \makecell{0.198 \\ ($\pm 0.329 $)}  & \makecell{ $\bm{0.189}$ \\ ($\pm\bm{ 0.904 }$)}& \makecell{$-0.179$ \\ ($\pm$0.177)}& \makecell{$\bm{-0.341}$ \\($\bm{\pm 0.620}$)} \\
         DNN w. BN & \makecell{ $\bm{0.127}$\\($\bm{\pm 2.778}$)}& \makecell{$\bm{0.193}$\\($\bm{\pm 2.412 }$)} & \makecell{0.205\\($\pm 18.817$)} & \makecell{$\bm{2.700}$ \\($\bm{\pm 0.544$})} &\makecell{$-10.462$ \\($\pm$9.105)} \\
         \hline
            & \multicolumn{5}{c}{Exp. 2}    \\
            \cline{2-6}
      DNN w. dropout & \makecell{ $\bm{0.119}$ \\ ($\bm{\pm 1.797}$)} & \makecell{ 0.202 \\ ($\pm 0.687$)}  &\makecell{$2.334$ \\ ($\pm 15.298$)}& \makecell{$\bm{3.841}$ \\($\bm{\pm 0.645}$)}& \makecell{$\bm{-13.794}$ \\ ($\bm{\pm 7.457)}$} \\
         DNN w. BN & \makecell{0.156\\($\pm 9.133$)}&  \makecell{ $\bm{0.196}$\\($\bm{ \pm 0.823}$)} &\makecell{$\bm{ 0.598}$ \\($\bm{\pm 89.937$})} & \makecell{$-5.098$\\ ($\pm 2.558$)} &\makecell{$ -191.78$\\ ($\pm$43.838)} \\
         \hline
            & \multicolumn{5}{c}{Exp. 3 }    \\
            \cline{2-6}
       DNN w. dropout & \makecell{$\bm{0.122}$\\($\bm{\pm 2.632}$)} & \makecell{2.122 \\($\pm 1.185 $)}  & \makecell{$\bm{0.170}$ \\($\bm{\pm 5.285}$)}&\makecell{$\bm{4.248}$ \\($\bm{\pm0.736}$)} & \makecell{$\bm{-13.670}$ \\($\bm{\pm 3.527}$)} \\
         DNN w. BN & \makecell{$0.138$\\($\pm 18.870$)}&  \makecell{$\bm{0.206}$\\($\bm{\pm 0.715}$)} & \makecell{0.271\\($\pm 134.062$)} &\makecell{$1.160$ \\($\pm 5.526$)} & \makecell{$-81.173$ \\($\pm$89.469)} \\
         \hline
                     & \multicolumn{5}{c}{Exp. 4}    \\
                     \cline{2-6}
      DNN w. dropout & \makecell{$\bm{0.113}$ \\($\bm{\pm 1.315}$)}& \makecell{0.194 \\($\pm 1.385$)} & \makecell{$\bm{0.232}$ \\($\bm{\pm 8.753}$) }& \makecell{$\bm{3.443}$ \\($\bm{\pm 0.764}$)} & \makecell{$\bm{-6.058}$ \\($\bm{\pm 3.992}$)}\\
        DNN w. BN & \makecell{$0.129$\\($\pm 5.911 $)}&  \makecell{$\bm{0.185}$\\($\bm{\pm 0.121}$)} & \makecell{0.395\\($\pm 49.916$)} & \makecell{$1.003$ \\($\pm 1.816$)} & \makecell{$-80.260$ \\($\pm$22.767)}\\
            \bottomrule
    \end{tabular}%
\begin{tablenotes}[flushleft]\footnotesize
    \item[]
    Note: All the $\textrm{MSE}$ and $R^{2}$ values have been multiplied by a factor of $10^{-2}$ and all the s.d. values has been multiplied by a factor of $10^{-5}$.
    \end{tablenotes}
  \label{tab:fundamentals}%
  \end{table}%

\section{Conclusion}
\label{sec:5}
In this study, we explored hyperparameter optimization techniques used in
A stock return prediction by applying DNN-based predictors. The experiment was validated
by considering different settings for the datasets, periods, and regularization.
We found that technical indicators are robust to an overfitting
during the HPO procedure, showing positive $R_{IS}$ and $R_{OS}$ values over different time periods, whereas the fundamental indicators are prone to an overfitting to the in-sample data. To summarize, dropout layers can efficiently decrease the risk of an overfitting and increase the model generalizability. 

This system can be seen as a first step toward a better and
more fruitful integration of the recent developments in HPO techniques. Future efforts for improving
the current solution will be devoted to the design of a neural architecture for the fundamental data, which are robust to an overfitting. Fundamental data evidently reflect the fundamental values, which can
serve as useful predictors or provide complementary information for a stock return prediction. 
We expect the development to improve the prediction accuracy by
combining fundamental and technical indicators.

\bibliographystyle{unsrt}
%\bibliography{BibFile}

%\begin{thebibliography}{}
%   \input{HyperparamterOptimizationForStock.bbl}
%\end{thebibliography}

\end{document}